\documentclass[ ]{revtex4}
\usepackage{hyperref}

\begin{document}
\title{Spontaneous Breaking of the BRST Symmetry in the ABJM theory}
 
 \author { Mir Faizal } 
\email{ mirfaizalmir@gmail.com}
\affiliation{Department of Physics and Astronomy, \\  University of Waterloo,   Waterloo,\\
Ontario N2L 3G1, Canada}
 
 \author { Sudhaker Upadhyay }
 \email{ sudhakerupadhyay@gmail.com;  sudhaker@boson.bose.res.in}
\affiliation{ Departamento de Fısica Teorica, Instituto de Fısica, \\
UERJ - Universidade do Estado do Rio de Janeiro,\\
Rua Sao Francisco Xavier 524, 20550-013 \\ Maracana, Rio de Janeiro, Brasil }

\begin{abstract}
In this paper, we will analyse the ghost condensation in the ABJM theory. 
We will perform our analysis in ${\cal N}=1$ superspace. 
We show that   in the Delbourgo-Jarvis-Baulieu-Thierry-Mieg gauge   the spontaneous breaking of BRST symmetry can occur 
in the ABJM theory. This spontaneous breaking of BRST symmetry is caused by 
ghost-anti-ghost condensation.
We will also show that  in  the ABJM theory, the ghost-anti-ghost 
the condensates remains present  in the modified abelian gauge.   
 Thus, 
the  spontaneous breaking of BRST symmetry in ABJM theory can  even occur in the modified abelian gauge. 
\end{abstract}

\maketitle
\section{Introduction}
 According to the $AdS_4/CFT_3$ correspondence the 
 field theory dual to the eleven dimensional supergravity is a 
 superconformal field theory  with $\mathcal{N} = 8$ supersymmetry. This is because apart from 
 a constant closed 7-form on $S^7$,
$AdS_4 \times S^7 \sim [SO(2,3)/ SO (1, 3)]\times  [SO(8)/ SO(7)] \subset OSp(8|4)/[SO(1,3) \times SO(7)]$. 
This group 
  $OSp(8|4)$ gets realized as $\mathcal{N} = 8$ supersymmetry of the dual field theory. Furthermore, the field content 
  of this dual 
   superconformal field theory comprises of 
 eight gauge valued scalar fields and sixteen  physical fermions. As this theory only has 
sixteen on shell degrees of freedom, so,  the gauge fields  cannot 
have any contribution to the on shell degrees of freedom.
Thus, the gauge sector of this theory is represented by Chern-Simons type actions. 
A theory called the 
the Bagger-Lambert-Gustavsson (BLG) theory meets all these requirements  \cite{bl, g}. 

 The gauge symmetry in the BLG theory is generated by  a Lie $3$-algebra rather than a Lie algebra and $SO(4)$ is the only known  
 example of a Lie $3$-algebra.
 It is possible to decompose the gauge symmetry generated by $SO(4)$ into $SU (2) \times SU (2)$. 
 If we do that for the BLG theory, then its  gauge symmetry   is  generated by ordinary Lie algebras. The gauge sector of the 
 BLG theory is now represented by two 
 Chern-Simons theories with levels $\pm k$ and the 
matter fields  exist in the bi-fundamental representation.  The  BLG theory only represents two M2-branes because its
 the gauge symmetry  is generated by the gauge group 
 $SU(2)_k \times SU(2)_{-k}$. However, it has been possible to extend the gauge group 
 to  $U(N)_k \times U(N)_{-k}$, and the resultant theory is called Aharony-Bergman-Jafferis-Maldacena (ABJM) 
theory \cite{abjm}.    Even though, the ABJM theory only has $\mathcal{N} =6$ supersymmetry,  
this supersymmetry   gets enhanced to 
$\mathcal{N} =8$ supersymmetry 
 for Chern-Simons levels, $k = 1, 2$ \cite{abjm2}.   
Furthermore,  for two M2-branes ABJM theory coincides with the BLG theory and thus has  $\mathcal{N} =8$ supersymmetry.

   It may be noted that as the ABJM theory has gauge symmetry, it cannot be quantized without getting rid 
of these unphysical degrees of freedom. This 
can be done by fixing a gauge. The gauge fixing condition can be incorporated at a quantum level by 
adding  ghost and gauge fixing terms to the original classical Lagrangian. It is known that 
for  a gauge theory the new effective Lagrangian 
constructed as the sum of the original classical Lagrangian with the gauge fixing and the ghost terms, 
is invariant a new set of transformations called the Becchi-Rouet-Stora-Tyutin (BRST) transformations \cite{brst, brst1}.
 Recently, BRST symmetry has also been studied  in non-linear gauges also \cite{nlbrst,nlbrst1}.
The BRST symmetry for the ABJM theory has also been studied \cite{mir, 0mir1, 0mir2, 0mir4, sudd}.

The ghost-anti-ghost condensation in gauge theories has been throughly 
 studied \cite{kon, z1, z2, z3, z4}. 
Such condensation can lead to a spontaneous breaking of the BRST and the anti-BRST symmetries. 
In fact, in recent past such ghost-anti-ghost
condensation has been proposed  as
a mechanism of providing the masses of off-diagonal gluons and off-diagonal ghosts
in the Yang-Mills theory in the Maximally Abelian gauge \cite{sch, kon1}. This mechanism helps in providing evidences for the 
infrared Abelian dominance \cite{hooft}, thereby justifies the dual
superconductor picture \cite{nambu, mandal, poly} of QCD vacuum for explaining quark confinement \cite{kon1,kon2,kon3,kon4}. 
It may be noted that breaking of BRST symmetry has led to many interesting consequences \cite{kon, n0, n1, n2, n4}.
 Therefore, these give us sufficient motivations to study the 
 spontaneous breaking of the BRST and the anti-BRST symmetries in ABJM theory.
 
With these motivations, in this paper, we establish the
  Chern-Simons-matter theory with different labels $k$ and $-k$
 in  non-linear Delbourgo-Jarvis-Baulieu-Thierry-Mieg (DJBTM) gauge. Further, 
the quantum actions generated due to
these gauge conditions are shown to posses  supersymmetric
BRST and anti-BRST invariance. A novel feature of ABJM theory in non-linear DJBTM gauge is observed that the presence of
ghost-anti-ghost condensates in the  theory breaks  both the BRST and anti-BRST symmetries spontaneously. 
In accordance with Nambu-Goldstone theorem the Faddeev-Popov ghost  and anti-ghost fields of the ABJM theory are identified 
as Nambu-Goldstone particles.
Furthermore, we construct the effective potentials in case ABJM theory explicitly 
which confirm the appearance of ghost-anti-ghost condensation 
as well as spontaneous symmetry breaking in the theory. 
Finally, we will analyse the ghost-anti-ghost condensation in the ABJM theory in modified maximally abelian (MA) gauge.

\section{The   ABJM Theory in ${\cal N}=1$ Superspace}
In this section we will review the construction of   Lagrangian    for ABJM theory in $\mathcal{N} =1$
superspace formalism. The three dimensional $\mathcal{N} =1$ superspace is parameterized by three spacetime coordinates along with 
a two component anti-commutating parameter, $\theta^a$. Now $Q_a = \partial_a -(\gamma^\mu \partial_\mu)^b_a \theta_b$, is the 
generator of $\mathcal{N} =1$ supersymmetry. This generator of $\mathcal{N} =1$ supersymmetry commutes with a superspace derivative,  
$D_a = \partial_a + (\gamma^\mu \partial_\mu)^b_a \theta_b$. This superspace derivative can be used to construct the 
Lagrangian    for ABJM theory in $\mathcal{N} =1$
superspace formalism.
As the ABJM theory is a 
gauge theory with the gauge group $U(N)_{k}  \times U(N)_{-k}$,  we can write the  Lagrangian    for ABJM theory as
\begin{equation}
{ \mathcal{L}_c} =  \mathcal{L}_{M} + \mathcal{L}_{CS} - \tilde{\mathcal{L}}_{CS},
\end{equation} 
where   $\mathcal{L}_{CS}, \tilde{\mathcal{L}}_{CS} $ 
are  Chern-Simons Lagrangians, 
and $\mathcal{L}_{M}$ is Lagrangian    for the matter fields. 
These Chern-Simons Lagrangian  are defined by
 \begin{eqnarray}
 \mathcal{L}_{CS} &=& \frac{k}{2\pi} \int d^2 \,  \theta \, \, 
  Tr \left[  \Gamma^a         \Omega_a
\right], 
\nonumber \\
 \tilde{\mathcal{L}}_{CS} &=& \frac{k}{2\pi} \int d^2 \,  \theta \, \, 
  Tr \left[  \tilde{\Gamma}^a         \tilde{\Omega}_a
\right], 
\end{eqnarray}
where $k$ is an integer  and 
 \begin{eqnarray}
 \Omega_a & = &  \frac{1}{2} D^b D_a \Gamma_b - \frac{i}{2} 
 [\Gamma^b , D_b \Gamma_a]    -
 \frac{1}{6} [ \Gamma^b ,
\{ \Gamma_b , \Gamma_a\}    ]  - \frac{1}{6}[\Gamma^b, \Gamma_{ab}],    \\
 \Gamma_{ab} & = & -\frac{i}{2} [ D_{(a}\Gamma_{b)} 
- i\{\Gamma_a, \Gamma_b\}    ],\nonumber \\
\tilde \Omega_a & = &\frac{1}{2} D^b D_a \tilde \Gamma_b 
- \frac{i}{2}  [\tilde \Gamma^b , D_b \tilde\Gamma_a]    -
 \frac{1}{6} [ \tilde \Gamma^b ,
\{ \tilde \Gamma_b ,  \tilde \Gamma_a\} ]  - \frac{1}{6}
[\tilde \Gamma^b, \tilde \Gamma_{ab}],    \\
 \tilde \Gamma_{ab} & = & -\frac{i}{2} [ D_{(a}\tilde \Gamma_{b)} 
- i\{\tilde \Gamma_a, \tilde \Gamma_b\}    ].
\end{eqnarray} 

The fields  $\Gamma_a$ and $\tilde \Gamma_a$
are   matrix valued spinor superfields suitable contracted with generator $T_A$
of   Lie algebra as $\Gamma_a= \Gamma_a^A T_A$ and $\tilde\Gamma_a= \tilde\Gamma_a^A T_A$, 
respectively and they are expressed in component form as 
\begin{eqnarray}
 \Gamma_a = \chi_a + B \theta_a + \frac{1}{2}(\gamma^\mu)_a A_\mu + i\theta^2 \left[\lambda_a -
 \frac{1}{2}(\gamma^\mu \partial_\mu \chi)_a\right], \nonumber \\
 \tilde\Gamma_a = \tilde\chi_a + \tilde B \theta_a + \frac{1}{2}(\gamma^\mu)_a \tilde A_\mu + i\theta^2 \left[\tilde \lambda_a -
 \frac{1}{2}(\gamma^\mu \partial_\mu \tilde\chi)_a\right]. 
 \end{eqnarray}
 Thus, in component form these Lagrangian are given by 
\begin{eqnarray}
  \mathcal{L}_{cs} &=& \frac{k}{4\pi}
\left( 2  \epsilon^{\mu \nu \rho} A_\mu     \partial_\nu  A_\rho 
 + \frac{4i}{3} A_\mu     A_\nu      A_\rho     + 
 E^a     E_a + \mathcal{D}_\mu     ( \chi^a(\gamma^\mu)_a^b      E_b)\right),
\nonumber \\
   \tilde {\mathcal{L}_{cs}} &=& \frac{k}{4\pi}
\left( 2 \epsilon^{\mu \nu \rho}  \tilde A_\mu     \partial_\nu   \tilde 
A_\rho 
 + \frac{4i}{3}  \tilde A_\mu      \tilde A_\nu       \tilde A_\rho  
 + 
  \tilde E^a      \tilde E_a +  \tilde{\mathcal{D}}_\mu     
(  \tilde \chi^a(\gamma^\mu)_a^b       \tilde E_b)\right).
\end{eqnarray}
The explicit expression for Lagrangian    of the matter fields  is given by 
\begin{eqnarray}
 \mathcal{L}_{M} &=& \frac{1}{4} \int d^2 \,  \theta \, \,  
Tr \left[  \nabla^a_{}         X^{I \dagger}               
\nabla_{a 
}               X_I     + 
  \mathcal{V}_{        } \right],
\end{eqnarray}
where the super-covariant derivatives for matrix valued complex scalar superfields
$ X^I$ and $X^{I  \dagger}$ are defined by  
\begin{eqnarray}
 \nabla_{a}              X^{I } &=& D_a  X^{I } + i \Gamma_a          
    X^I - i  X^I        \tilde\Gamma_a      , \nonumber \\ 
 \nabla_{a}              X^{I \dagger} &=& D_a  X^{I  \dagger} 
- i X^{I  \dagger}       \Gamma_a    
        + i \tilde\Gamma_a            X^{I  \dagger}, 
\end{eqnarray}
and $\mathcal{V}      $ is the potential term  given by 
\begin{eqnarray}
 \mathcal{V}      & =& \frac{16\pi}{k}\epsilon^{IJ} \epsilon_{KL} 
[ X_I       X^{K \dagger}        X_J       X^{L\dagger}]. 
\end{eqnarray}
The classical Lagrangian    for ABJM theory ${\cal L}_c$  remains invariant  under the  following  gauge 
transformation 
\begin{eqnarray}
 \delta \Gamma_a =  \nabla_a {     } \Lambda, 
&&   \delta \tilde\Gamma_a = \tilde\nabla_a {     }
 \tilde\Lambda, \nonumber \\ 
\delta X^{I } = i(\Lambda{     } X^{I }  - X^{I }{     }\tilde \Lambda ),  
&&  \delta  X^{I \dagger  } 
= i(   \tilde \Lambda {     } X^{I\dagger  }-X^{I\dagger  }{     } \Lambda), 
\end{eqnarray}
where $ \Lambda = \Lambda^A T_A$ and $\tilde \Lambda = \tilde \Lambda^A\tilde  T_A$ are parameters of transformations. 
The Lagrangian    for the ABJM theory is invariant under these  gauge transformations. 
\section{ ABJM Theory in DJBTM Gauge}
The gauge invariance of ABJM theory reflects that the theory is endowed 
with some spurious degrees of freedom.
In order to quantize the theory correctly we need to fix the gauge. 
In this section, we will analyse the ABJM theory in Delbourgo-Jarvis-Baulieu-Thierry-Mieg  (DJBTM) Gauge \cite{1z, 2z}. 
We start  with  proper choices of
the covariant   gauge fixing conditions for ABJM theory 
which remove the spurious degrees of freedom as follows 
$
G_1 \equiv D^a \Gamma_a  =0,\,  \tilde G_1 \equiv D^a \tilde{\Gamma}_a =0
$ \cite{mir}.
These  gauge fixing conditions can be incorporated at the quantum level by adding the following  gauge fixing term with gauge parameter $\alpha$  to 
the original Lagrangian   ,
\begin{equation}
\mathcal{L}_{gf} = \int d^2 \,  \theta \, \,  \mbox{Tr}   \left[ b   (D^a \Gamma_a) + \frac{\alpha}{2}b   b -
  \tilde{b}    (D^a \tilde{\Gamma}_a) - \frac{\alpha}{2}\tilde{b}    \tilde b 
\right],
\end{equation}
where $b$ and $\tilde b$ are Nakanishi-Lautrup type auxiliary fields.
The Faddeev-Popov ghost terms corresponding to the above gauge fixing term can be written
in terms of ghost fields $c, \tilde{c}$ and corresponding anti-ghost fields $\bar c,  \tilde{\bar{c}}$
explicitly as  
\begin{equation}
\mathcal{L}_{gh} = i\int d^2 \,  \theta \, \,    \mbox{Tr} 
[  \bar{c}    D^a \nabla_a    c - \tilde{\bar{c}}    D^a \tilde{\nabla}_a   \tilde{c} ].
\end{equation}
These gauge-fixing and ghost terms are BRST exact and  defined together by
\begin{equation}
{\cal L}_{g}={\cal L}_{gf} + {\cal L}_{gh}. 
  \end{equation}
Now, the nilpotent  BRST transformations (i.e. $s_b^2=0$) are constructed by
\begin{eqnarray}
s_b \,\Gamma_{a} = \nabla_a    c, && s_b \, \tilde\Gamma_{a} =\tilde\nabla_a    
 \tilde c, \nonumber \\
s_b  \,c = -  \frac{1}{2} {[c,c]}_ { }, && s_b  \,\tilde{ {c}} = -  \frac{1}{2} [\tilde{ {c}} ,  \tilde c]_{ }, \nonumber \\
s_b  \,\bar{c} = ib, && s_b  \,\tilde {\bar c} = i\tilde b, \nonumber \\ 
s_b  \,b =0, &&s_b  \, \tilde b= 0, \nonumber \\ 
s_b  \, X^{I } = i c   X^{I } -  iX^{I }  \tilde c, 
 &&  s_b  \, X^{I \dagger }
 = i   \tilde c    X^{I \dagger } - i  X^{I \dagger }  c, 
\nonumber \\
s_b  \, Y^{I } = i  \tilde c   Y^{I } -iY^{I }   c,  &&  
s_b  \, Y^{I \dagger } = i c   Y^{I \dagger } -  i Y^{I \dagger }
  \tilde c,\label{brs}
\end{eqnarray}
which leave  the effective ABJM Lagrangian      $ {\cal L}_{ABJ}={\cal L}_c + {\cal L}_{g} 
  $  invariant.
  With the help of this BRST symmetry the gauge fixing and ghost parts of the 
  effective Lagrangian    can be expressed as 
  \begin{eqnarray}
  {\cal L}_g =is_b\int d^2 \,  \theta \, \,    \mbox{Tr} \left[  \tilde{\bar c}  D^a \tilde\Gamma_a +\frac{\alpha}{2}\tilde{b} - \bar c  D^a \Gamma_a -  \frac{\alpha}{2} {b} 
 \right].
  \end{eqnarray}
Furthermore, we explore the another nilpotent symmetry so-called anti-BRST transformations for ABJM theory, where the role of
ghost and anti-ghost fields are interchanged, as follows
\begin{eqnarray}
s_{ab} \,\Gamma_{a} = \nabla_a   \bar c, && s_{ab} \, \tilde\Gamma_{a} =\tilde\nabla_a    
 \tilde{\bar c}, \nonumber \\
s_{ab}  \,\bar c = -  \frac{1}{2} {[\bar c, \bar c]}_ { }, && s_{ab}  \,\tilde{ \bar {c}} = -  \frac{1}{2} [\tilde{\bar  {c}} ,  \tilde{\bar c}]_{ }, \nonumber \\
s_{ab}  \, {c} = i\bar b, && s_{ab}  \,\tilde { c} = i\tilde {\bar b}, \nonumber \\ 
s_{ab}  \,\bar b =0, &&s_{ab}  \, \tilde {\bar b}= 0, \nonumber \\ 
s_{ab}  \, X^{I } = i \bar c   X^{I } -  iX^{I }  \tilde {\bar c}, 
 &&  s_{ab}  \, X^{I \dagger }
 = i   \tilde {\bar c}    X^{I \dagger } - i  X^{I \dagger }  \bar c, 
\nonumber \\
s_{ab}  \, Y^{I } = i  \tilde {\bar c}   Y^{I } -iY^{I }  \bar c,  &&  
s_{ab}  \, Y^{I \dagger } = i\bar c   Y^{I \dagger } -  i Y^{I \dagger }
  \tilde {\bar c}.\label{antibrs}
\end{eqnarray}
Here we remark that the newly added auxiliary fields $\bar b$ and $\tilde{\bar b}$ can be 
expressed in terms of original auxiliary fields $b$ and $\tilde b $ as following:
\begin{eqnarray}
\bar b =-b+i[c,\bar c],\ \ \tilde{\bar b} =-\tilde b +i[\tilde c,
\tilde{\bar c}].\label{cf}
\end{eqnarray}
These conditions are similar to Curci-Ferrari (CF) type restriction.
 
Now,  we will construct the effective Lagrangian for the ABJM theory
in   DJBTM   gauge.
 For this purpose, we construct the sum of 
  gauge-fixing and ghost parts of the effective Lagrangian    
 in  DJBTM  gauge  which is defined by
\begin{eqnarray}
{\cal L}^{DJ}_g 
&=&\int d^2\theta\    \mbox{Tr} \left[ b   (D^a \Gamma_a) + \frac{\alpha}{2}b   b -
  \tilde{b}    (D^a \tilde{\Gamma}_a) - \frac{\alpha}{2}\tilde{b}    \tilde b 
  + i\bar{c}    D^a \nabla_a    c\right.\nonumber\\ 
  &&\left. -i \tilde{\bar{c}}    D^a \tilde{\nabla}_a   \tilde{c} -\frac{\alpha}{2} i[c,\bar c]b +\frac{\alpha}{8}[\bar c, \bar c][c, c]
 + \frac{\alpha}{2} i[\tilde c, \tilde{\bar c}]\tilde b \right.\nonumber\\ 
  &&\left. -\frac{\alpha}{8}[
 \tilde{\bar c}, \tilde{\bar c}][\tilde c, \tilde c] \right]. 
 \end{eqnarray}
 This can further be written as
 \begin{eqnarray}
{\cal L}^{DJ}_g 
 &=&\int d^2\theta\    \mbox{Tr} \left[ b   (D^a \Gamma_a) + \frac{\alpha}{2}b   b -
  \tilde{b}    (D^a \tilde{\Gamma}_a) - \frac{\alpha}{2}\tilde{b}    \tilde b 
  + i\bar{c}    D^a \nabla_a    c \right.\nonumber\\
  &&\left. -i \tilde{\bar{c}}    D^a \tilde{\nabla}_a   \tilde{c}
  - \frac{\alpha}{2} i[c,\bar c]b -\frac{\alpha}{4}[ c, \bar c][c, \bar c]
 + \frac{\alpha}{2} i[\tilde c, \tilde{\bar c}]\tilde b \right.\nonumber\\
  &&\left.+\frac{\alpha}{4}[
 \tilde{  c}, \tilde{\bar c}][\tilde c, \tilde{\bar c}] \right].
\end{eqnarray}
The effective Lagrangian    for ABJM theory in DJBTM  gauge,   ${\cal L}_c + {\cal L}^{DJ}_{g}$, is also
invariant under above set of BRST and anti-BRST transformations mentioned
  in Eqs. (\ref{brs}) and (\ref{antibrs}).
Further, we can express the Lagrangian    ${\cal L}^{DJ}_g$ in terms
of both BRST as well as anti-BRST exact term
\begin{eqnarray}
{\cal L}^{DJ}_g &=&\frac{i}{2}s\bar s\int d^2\theta\   \mbox{Tr} [\Gamma_a \Gamma^a -\tilde \Gamma_a\tilde\Gamma^a -i\alpha
\bar c c +i\alpha \tilde{\bar c}\tilde c].
\end{eqnarray}
Now, to inspect the non-zero gauge 
parameter,   we   write the above Lagrangian     as 
\begin{eqnarray}
{\cal L}^{DJ}_g &=&\int d^2\theta\    \mbox{Tr} \left[ \frac{\alpha}{2}\left(b-\frac{1}{2}i[c, \bar c]  +\frac{1}{\alpha} D_a\Gamma^a \right)^2 -\frac{1}{2\alpha}(D_a\Gamma^a)^2 
  + i\bar{c}    D^a \nabla_a    c \right.\nonumber\\
  &&-\left.  \frac{\alpha}{2}\left(\tilde b-\frac{1}{2}i[\tilde c, \tilde{\bar c}]  +\frac{1}{\alpha} D_a\tilde \Gamma^a \right)^2 +\frac{1}{2\alpha}(D_a\tilde \Gamma^a)^2 
   -i \tilde{\bar{c}}    D^a \tilde{\nabla}_a   \tilde{c}\right.\nonumber\\
  && -\left.  \frac{\alpha}{8}[ c, \bar c][c, \bar c]
 +\frac{\alpha}{8}[
 \tilde{  c}, \tilde{\bar c}][\tilde c, \tilde{\bar c}] \right].
\end{eqnarray}
For analysing the  spontaneous breaking of BRST symmetry,  the non-linear auxiliary 
fields $b$ and $\tilde b$ 
could play an important 
role as an order
parameters for the BRST symmetry breaking. Therefore,
we would not remove them by
using equations of motion.
\section{Spontaneous Breaking of BRST  Symmetry }
In this section, we 
describe the breaking of BRST 
supersymmetry spontaneously
in case of ABJM theory.
To do so, let us
 define the potential $V(b, \tilde b)$ for multiplier
 fields 
 $b$ and $\tilde{b}$ such that
 \begin{eqnarray}
 V(b,\tilde{b}) &=& \int d^2\theta\    \mbox{Tr} \left[- \frac{\alpha}{2}\left(b-\frac{1}{2}i[c, \bar c] 
 +\frac{1}{\alpha} D_a\Gamma^a \right)^2  \right. \nonumber \\ && \left.
 + \frac{\alpha}{2}\left(\tilde b-\frac{1}{2}i[\tilde c, \tilde{\bar c}]  +\frac{1}{\alpha} D_a\tilde \Gamma^a \right)^2\right]. 
 \end{eqnarray}
The potential has its extremum for gauge parameter $\alpha$ for any integer value either at
\begin{eqnarray}
b=\frac{1}{2}i[c, \bar c]  -\frac{1}{\alpha} D_a\Gamma^a\  \mbox{and}\ \
\tilde b= \frac{1}{2}i[\tilde c, \tilde{\bar c}]  -\frac{1}{\alpha} D_a\tilde \Gamma^a.
\end{eqnarray}
The  vacuum states 
of non-linear bosonic fields $b$ and $\tilde b$ are given by
\begin{equation}
\langle 0| b|0\rangle =\frac{1}{2}\langle 0|i[c,\bar c]|0\rangle,\ \
\langle 0|\tilde b|0\rangle =\frac{1}{2}\langle 0|i[\tilde c, \tilde{\bar c}]|0\rangle,
\end{equation}
where we have utilized the  invariance such that
$\langle \Gamma_a\rangle =0$ and $\langle \tilde\Gamma_a\rangle =0$.  
In case of ghost-anti-ghost condensations appear
such that
\begin{eqnarray}
   \langle 0|i[c,\bar c]|0\rangle \neq 0,\ \
   \langle 0|i[\tilde c, \tilde{\bar c}]|0\rangle\neq 0,
\end{eqnarray}
 the non-linear fields $b$ and $\tilde b$
acquire non-vanishing vacuum to vacuum expectation values (VEVs), i.e., 
\begin{eqnarray}
\langle 0| b|0\rangle =\frac{1}{2}\langle 0|i[c,\bar c]|0\rangle \neq 0,\ \
\langle 0|\tilde b|0\rangle =\frac{1}{2}\langle 0|i[\tilde c, \tilde{\bar c}]|0\rangle\neq 0.
\end{eqnarray}
Consequently, these non-vanishing VEVs  break the BRST symmetry spontaneously
as follows:
\begin{eqnarray}
\langle 0| s_b \bar c|0\rangle = \langle 0|i b|0\rangle =-\frac{1}{2}\langle 0| [c,\bar c]|0\rangle \neq 0,\nonumber\\
\langle 0| s_b \tilde {\bar c}|0\rangle = \langle 0|i\tilde b|0\rangle =-\frac{1}{2}\langle 0| [\tilde c, \tilde{\bar c}]|0\rangle\neq 0.
\end{eqnarray}
Using CF conditions given in Eq. (\ref{cf}), it is easy to see
that spontaneous breaking of anti-BRST symmetry also occurs in this case
\begin{eqnarray}
\langle 0| s_{ab}  c|0\rangle = \langle 0|i \bar b|0\rangle =-\frac{1}{2}\langle 0| [c,\bar c]|0\rangle \neq 0,\nonumber\\
\langle 0| s_{ab }\tilde {c}|0\rangle = \langle 0|i\tilde {\bar b}|0\rangle =-\frac{1}{2}\langle 0| [\tilde c, \tilde{\bar c}]|0\rangle\neq 0.
\end{eqnarray}
According to Nambu-Goldstone theorem we note that,
corresponding to these spontaneous symmetries breaking there exist
massless  Nambu-Goldstone particles.
For instance, these ghosts and anti-ghosts can be identified as
Nambu-Goldstone particles.

To determine whether such ghost-anti-ghost condensation as well as
spontaneous symmetry breaking take place or not, we express the
effective potential in case of ABJM theory as
\begin{eqnarray}
V(b,\tilde{b}, \phi, \tilde{\phi}) =
V(\phi,\tilde \phi)+\int d^2\theta\    \mbox{Tr} \left[ -\frac{\alpha}{2} \left( b +\frac{1}{2\alpha} \phi\right)^2
+\frac{\alpha}{2} \left(\tilde b +\frac{1}{2\alpha} \tilde \phi\right)^2\right],
\end{eqnarray}
where  $\phi \sim -\alpha \langle 0|i[c, \bar c] |0\rangle$ and $\tilde\phi \sim -\alpha \langle 0|i[\tilde c, \tilde{\bar c}] |0\rangle$. 
However, it can be noticed that   
for Landau gauge condition where gauge parameter takes zero value such kind of
ghost-anti-ghost condensation
and  consequently 
spontaneous symmetry breaking  does not appear for ABJM theory
in ${\cal N}=1$ superspace.
This result verifies
 the conventional  use
of the BRST symmetry for ABJM theory in the linear gauge. 
\section{ABJM Theory in MA Gauge  }
Now we analyse ghost-anti-ghost condensation in modified maximally abelian (MA) gauge  \cite{4z, 5z}. Thus, we start by 
decomposing  the gauge fields in diagonal and off-diagonal components as
follows
 \begin{eqnarray}
 \Gamma_a= \gamma_a^iT_i +  {\Upsilon}^\alpha_a T_\alpha,\ \  
 \tilde\Gamma_a= \tilde\gamma_a^i \tilde T_i +  \tilde{\Upsilon}^\alpha_a \tilde T_\alpha,  
\end{eqnarray}
where $T_i\in \mathcal{\cal H}$ and $T_\alpha\in  U(N)_k -\mathcal{\cal H}$
with $\mathcal{\cal H}$ being the Cartan subalgebra of the Lie algebra $U(N)_k$.
Similarly, $\tilde T_i\in \mathcal{\cal H}$ and $\tilde T_\alpha\in U(N)_{-k}-\mathcal{\cal H}$
with $\mathcal{\cal H}$ being the Cartan subalgebra of the Lie algebra $U(N)_{-k}$.
Now, the Lagrangian    in MA gauge   is constructed in terms of BRST exact quantity
\begin{eqnarray}
{\cal L}_g^{MA} &=&-is\int d^2\theta\    \mbox{Tr}\left[\bar c\left\lbrace \nabla_a[\gamma]\Upsilon^a 
+\frac{\alpha}{2}b\right\rbrace -i\frac{\zeta}{2}\bar c[\bar{c}, c] -i\frac{\zeta}{4} c[\bar{c},\bar{c}] \right.\nonumber\\
&-&\left.\tilde{\bar c}\left\lbrace \tilde\nabla_a[\tilde\gamma]\tilde\Upsilon^a +\frac{\alpha}{2}\tilde b\right\rbrace
+i\frac{\zeta}{2}\tilde{\bar c}[\tilde{\bar{c}}, \tilde c] +i\frac{\zeta}{4} \tilde c[\tilde{\bar{c}}, \tilde{\bar c}]\right].
\end{eqnarray}
Utilizing BRST transformation the above Lagrangian    is further expanded as
\begin{eqnarray}
{\cal L}_g^{MA} &=&\int d^2\theta\    \mbox{Tr}\left[b\nabla_a [\gamma]\Upsilon^a
+ \frac{\alpha}{2}b^2\right.\nonumber\\
&&\left. +i\bar c\nabla_a[\gamma]\nabla^a [\gamma]c-i[\bar c, \Upsilon_a] [c,\Upsilon^a]+i\bar c\nabla_a[\gamma]( [\Upsilon^a, c] )\right.\nonumber\\
&&\left.+ i\bar c[\nabla_a[\gamma]\Upsilon^a, c] +\frac{\zeta}{8}[\bar c, \bar c][c,c] +\frac{\zeta}{4}[\bar c, \bar c][c,c] 
\right.\nonumber\\
&&\left. +i\frac{\zeta}{2}c[b,\bar c]-i\zeta b [\bar c,c] +\frac{\zeta}{4}[\bar c,\bar c][c, c]\right.\nonumber\\
&& \left. -\tilde b\tilde \nabla_a [\tilde\gamma]\tilde\Upsilon^a
- \frac{\alpha}{2}\tilde b^2-i\tilde{\bar c}\tilde\nabla_a[\tilde\gamma]\tilde\nabla^a [\tilde\gamma]\tilde c\right.\nonumber\\
&&\left. +i[\tilde{\bar c}, \tilde\Upsilon_a] [\tilde c,
\tilde\Upsilon^a]-i\tilde{\bar c}\tilde\nabla_a[\tilde\gamma]( [\tilde\Upsilon^a, 
\tilde c] )\right.\nonumber\\
&&\left. -i\tilde{\bar c}[\tilde\nabla_a[\tilde\gamma]\tilde \Upsilon^a, \tilde c] 
-\frac{\zeta}{8}[\tilde{\bar c}, \tilde{\bar c}][\tilde c, \tilde c] \right.\nonumber\\
&&\left. -\frac{\zeta}{4}[\tilde{\bar c}, 
\tilde{\bar c}][\tilde c, \tilde c] -i\frac{\zeta}{2}\tilde c[
\tilde b, \tilde{\bar c}]+i\zeta b [\tilde{\bar c}, \tilde c] \right.\nonumber\\
&&\left.-\frac{\zeta}{4}[
\tilde{\bar c}, \tilde{\bar c}][\tilde c, \tilde c]
\right].
\end{eqnarray}
In the modified MA gauge , the requirement of the orthosymplectic 
invariance yields the quartic ghost interaction as $\zeta =\alpha$. In the modified MA gauge , the above expression reduces to
\begin{eqnarray}
{\cal L}_g^{MMA} &=&\int d^2\theta\    \mbox{Tr}\left[\frac{\alpha}{2}\left(b- i[\bar c, c] 
+\frac{1}{\alpha} \nabla_a[\gamma] \Upsilon^a\right)^2  -\frac{1}{2\alpha} (\nabla_a[\gamma] \Upsilon^a )^2 
\right.\nonumber\\
&&\left. - i[\bar c, c] [\Upsilon_a,\Upsilon^a] + i\bar c\nabla_a[\gamma]( [\Upsilon^a, c] ) 
-\frac{\alpha}{2}\left(\tilde b- i[\tilde{\bar c}, \tilde c] +\frac{1}{\alpha}\tilde \nabla_a[\tilde\gamma] \tilde\Upsilon^a\right)^2 
  \right.\nonumber\\
&&\left.+i\tilde{\bar c}\tilde\nabla_a[\tilde\gamma]\tilde\nabla^a [\tilde\gamma]c 
+i[\tilde{\bar c}, \tilde c] [\tilde\Upsilon_a,\tilde\Upsilon^a]-i\tilde{\bar c}\tilde\nabla_a[\tilde\gamma]
( [\tilde\Upsilon^a, \tilde c] )  
\right.\nonumber\\
&&\left.  -i\bar c\nabla_a[\gamma]\nabla^a [\gamma]c +\frac{1}{2\alpha} (\tilde\nabla_a[\tilde\gamma] \tilde\Upsilon^a )^2 
\right].
\end{eqnarray}
The potential for non-linear field $b$ and $\tilde{b}$ has extremum either at
\begin{eqnarray}
b&=& i[\bar c, c] -\frac{1}{\alpha} \nabla_a[\gamma] \Upsilon^a, \nonumber \\  
\tilde b&=& i[\tilde{\bar c}, \tilde c] -\frac{1}{\alpha}\tilde \nabla_a[\tilde\gamma] \tilde\Upsilon^a.
\end{eqnarray}
So, the  vacuum is defined as
\begin{eqnarray}
\langle 0|b|0\rangle &=& \langle 0|i[\bar c, c]|0\rangle -\frac{1}{\alpha} \langle 0| \nabla_a[\gamma] 
\Upsilon^a|0\rangle,   
\nonumber \\ 
 \langle 0|\tilde b|0\rangle &=& \langle 0| i[\tilde{\bar c}, \tilde c]|0\rangle -\frac{1}{\alpha} \langle 0|\tilde \nabla_a[\tilde\gamma] 
 \tilde\Upsilon^a|0\rangle.
\end{eqnarray}
This shows that even  in modified MA gauge  the ghost-anti-ghost condensates remains present in the ABJM theory and due to which the 
the spontaneous breaking of the BRST symmetry occurs.
An advantage of the spontaneous BRST supersymmetry breaking is that 
the Nambu-Goldstone particle  associated with the spontaneous breaking of 
the BRST symmetry or the  spontaneous breaking of 
anti-BRST  symmetry can be  identified with the 
diagonal anti-ghost  or diagonal ghost, respectively. Thus, the 
diagonal ghost and the diagonal anti-ghost are massless. This result is consistent with 
infrared Abelian dominance. As  infrared Abelian dominance  is expected 
to be realized, if the off-diagonal components of ghosts become 
massive while the diagonal components remain massless. 

\section{Conclusion}
In this paper, we analysed the ABJM theory in Delbourgo-Jarvis and Baulieu-Thierry-Mieg (DJBTM) gauge  and 
modified maximally abelian (MA) gauge. Furthermore, 
we have investigated the quantum actions for the theory admitting supersymmetric
BRST invariance. We have observed that due to 
ghost-anti-ghost condensates appear in non-linear DJBTM gauge
the non-linear bosonic fields admit non-vanishing 
vacuum expectation values (VEVs). 
Consequently, a spontaneous breaking of the supersymmetric BRST invariance 
has occurred in the theory. 
We also demonstrated that even  in modified MA gauge,
the ghost-anti-ghost condensates remains present in the ABJM theory, and due to which the 
the  spontaneous breaking of the BRST symmetry occurs.
We have identified the ghost  and anti-ghost fields present in the theory as 
a Nambu-Goldstone particles according to Nambu-Goldstone theorem.
To confirm the appearance of ghost-anti-ghost condensation 
as well as spontaneous symmetry breaking we constructed  an 
effective potential for the ABJM theory.

It may be noted just as strings can end on D-branes in string theory, M2-branes can also end on 
 M9-branes, M5-branes  and gravitational waves in M-theory
  \cite{ber}. 
  Boundary conditions for open M2-branes in presence of a flux have also been studied \cite{ber1}. 
  A system of multiple M2-branes ending on a M5-brane can be used to learn about the physics of 
  M5-branes. Thus,  a system a system of M2-branes ending on a M5-brane with a constant 
  $C$ field in the background has been used to motivate the study of a novel quantum geometry 
 on M5-brane world-volume \cite{ber2}. In fact, the BLG theory with a Nambu-Poisson 3-bracket has been 
 identified with the M5-brane action in presence of a large $C$ field \cite{ber4}. 
 The action for M2-branes in presence of a boundary has been constructed in $\mathcal{N} =1$ superspace formalism 
 \cite{0mir2, mir9, mir8, 1ber}. This theory is made gauge invariant by adding extra boundary degrees of freedom such that 
 the gauge transformation of the boundary theory exactly cancels the boundary piece generated by gauge transformation of 
 the bulk theory. Similarly, the BRST transformation of the boundary  theory  exactly cancels by boundary term generated by 
 the BRST transformation of the bulk theory. It will be interesting to investigate what happens to this system, if the 
 BRST symmetry is broken on the boundary or in the bulk due to the ghost-anti-ghost condensation.

\end{document}